\documentstyle [12pt]{article}
\begin{document}
\title{ Scale Symmetry Spontaneously Broken by Asymptotic Behavior 
\protect\\  } \author{E.I. Guendelman  
\\{\it Physics Department, Ben-Gurion University, Beer-Sheva
84105, Israel}}

\maketitle
\bigskip

\begin{abstract}
Conserved quantities are obtained and analyzed in the new models with global
scale invariance recently proposed. Such models allow for non trivial 
scalar field potentials and masses for particles, so that the scale symmetry 
must be broken somehow. We get to this conclusion by showing that the 
infrared behavior of the conserved currents is singular so that there are no
conserved charges associated with the global scale symmetry. The scale symmetry 
plays nevertheless a crucial role in determining the structure of the theory
and it implies that in some high field regions the potentials become flat.
 
\end{abstract}

\section{Introduction}

The idea of scale invariance, which implies that physics at all scales must be
the same is an appealing thought. A straightforward application of scale invariance
to the construction of physical theories leads however to consequences
that are very much against what we observe in the universe, since in physics
certain scales appear, through the appearance of typical atomic sizes, 
typical masses for particles, etc.

One can attempt to construct scale invariant theories where the scale 
invariance is spontaneously broken in the usual sense. This situation is
also not satisfactory, since then Goldstone's theorem$^{1}$ tell us that 
there must be a massless boson, associated with scale invariance$^{2}$, which 
is not observed so far.

It is known however that violent infrared behavior can invalidate Goldstone's
theorem in certain instances. For example the famous chiral U(1) problem in 
QCD is solved, in spite of the fact that even in the presence of the chiral 
U(1) anomaly, one can define a locally conserved chiral U(1) current. Local
current conservation nevertheless does not guarantee conservation of the 
integrated charge if the current has a singular infrared behavior, as it is
the case for QCD instantons. The absence of a conserved charge does not allow
us therefore to apply Goldstone's theorem and therefore the chiral U(1) 
problem in QCD is solved$^{3}$.

Here we will see that something like this can happen in the case of dilatation 
symmetry, although the effect is already evident at the classical level and there 
is no need to consider instantons. Indeed a model is found where dilatation
symmetry is exact, a local conservation law follows, but the infrared behavior
of the spatial components of the conserved dilatation current is so singular
in the infrared that charge can easily "leak out" to infinity, thus 
invalidating the possibility of a conserved scale charge. Thus we obtain,
non conservation of the scale charge without need to introduce any scale symmetry
breaking terms in the action.

The model is constructed by allowing, in addition to the usual volume element
used for integration in the action $\sqrt{-g}d^{4}x$, where 
$ g =  det g_{\mu\nu}$, 
another one$^{4}$,           
$\Phi d^{4}x$, where $\Phi$ is a density built out of degrees of freedom      
independent of that of $g_{\mu\nu}$. To achieve global scale invariance,      
also a "dilaton" $\phi$ has to be introduced$^{5}$.
        For example, given 4-scalars $\varphi_{a}$ (a =                       
1,2,3,4), one can construct the density                                       

\begin{equation}                                                             
\Phi =  \varepsilon^{\mu\nu\alpha\beta}  \varepsilon_{abcd}                   
\partial_{\mu} \varphi_{a} \partial_{\nu} \varphi_{b} \partial_{\alpha}       
\varphi_{c} \partial_{\beta} \varphi_{d}                                      
\end{equation}    

and we allow global scale transformations for the $\varphi_{a}$ to be 
independent to those of the metric $g_{\mu\nu}$. The infrared behavior of the 
$\varphi_{a}$ fields plays an essential role in the singular infrared
behavior of the dilatation current and in the global non conservation of
the dilatation charge, although there is a locally conserved dilatation 
current.

Although no dilatation charge exists, the dilatation symmetry has important 
consequences in the structure of the theory, the form of the scalar field 
potentials and their interactions to other fields, etc.

\section{The Model}
As mentioned in the introduction, we look at an action which uses both 
measures of integration $\sqrt{-g}d^{4}x$ and $\Phi  d^{4} x $ and consider
therefore the form,

\begin{equation}
S = \int L  d^{4} x
\end{equation} 

where
\begin{equation}                                                              
L =  L_{1} \Phi   +  L_{2} \sqrt{-g}               
\end{equation}                                                                

        One can notice that $\Phi$ is the total derivative of something,      
for example, one can write                                                        
\begin{equation}                                                              
\Phi = \partial_{\mu} ( \varepsilon^{\mu\nu\alpha\beta}                       
\varepsilon_{abcd} \varphi_{a}                                                
  \partial_{\nu} \varphi_{b}                                                  
\partial_{\alpha}                                                             
\varphi_{c} \partial_{\beta} \varphi_{d}).                                    
\end{equation}                                                                
                 
This means that there is a  shift symmetry that can be applied  on      
$L_{1}$                                                                       
$L_{1} \rightarrow L_{1}$ + constant.
Since such shift
just adds the integral of a total divergence to the action (2)-(3) and it does    
not affect therefore the equations of motion of the theory.                                                        
	In the action (2)-(3) the measure carries degrees of freedom
independent of that of the metric and that of the matter fields. The most
natural and successful formulation of the theory is achieved when the
connection coefficients are also treated as an independent degrees  of freedom.
 This is what is usually referred to as the first order formalism.

 Here $L_{1}$ and
$L_{2}$ are taken to be
$\varphi_{a}$  independent.

	There is a good reason not to consider mixing of  $\Phi$ and
$\sqrt{-g}$ , like
for example using
\begin{equation}
\frac{\Phi^{2}}{\sqrt{-g}} 
\end{equation}		 	

this is because (2)-(3) is invariant (up to the integral of a total
divergence) under the infinite dimensional symmetry
\begin{equation}
\varphi_{a} \rightarrow \varphi_{a}  +  f_{a} (L_{1})	
\end{equation}
where $f_{a} (L_{1})$ is an arbitrary function of $L_{1}$ if $L_{1}$ and
$L_{2}$ are $\varphi_{a}$
independent. Such symmetry (up to the integral of a total divergence) is
absent if mixed terms (like (5)) are present.  Therefore (2)-(3) is 
considered for the case when no dependence on the measure fields $\varphi_{a}$
appears in $L_{1}$ or $L_{2}$

\section{The Action Principle for a Scalar Field}

	We will study now the dynamics of a scalar field $\phi$ interacting
with gravity as given by the following action
\begin{equation}
S_{\phi} =  \int L_{1} \Phi d^{4} x  +  \int L_{2} \sqrt{-g}   d^{4} x
\end{equation}
\begin{equation}
L_{1} = \frac{-1}{\kappa} R(\Gamma, g) + \frac{1}{2} g^{\mu\nu}
\partial_{\mu} \phi \partial_{\nu} \phi - V(\phi) 
\end{equation}	
\begin{equation}
L_{2} = U(\phi)
\end{equation}
\begin{equation}	
R(\Gamma,g) =  g^{\mu\nu}  R_{\mu\nu} (\Gamma) , R_{\mu\nu}
(\Gamma) = R^{\lambda}_{\mu\nu\lambda}
\end{equation}
\begin{equation}
R^{\lambda}_{\mu\nu\sigma} (\Gamma) = \Gamma^{\lambda}_
{\mu\nu,\sigma} - \Gamma^{\lambda}_{\mu\sigma,\nu} +
\Gamma^{\lambda}_{\alpha\sigma}  \Gamma^{\alpha}_{\mu\nu} -
\Gamma^{\lambda}_{\alpha\nu} \Gamma^{\alpha}_{\mu\sigma}.	 
\end{equation}

 In the action the measure carries degrees of freedom              
independent of that of the metric and that of the matter fields. The most     
natural and successful formulation of the theory is achieved when the         
connection coefficients are also treated as an independent degrees  of freedom
This is what is usually referred to as the first order formalism.     
Therefore, in the variational principle $\Gamma^{\lambda}_{\mu\nu},
g_{\mu\nu}$, the measure fields scalars
$\varphi_{a}$ and the  scalar field $\phi$ are all to be treated
as independent
variables although the variational principle may result in equations that
allow us to solve some of these variables in terms of others.

\section{Global Scale Invariance}

	If we perform the global scale transformation ($\theta$ =
constant) 
\begin{equation}
g_{\mu\nu}  \rightarrow   e^{\theta}  g_{\mu\nu}	
\end{equation}
then (9) is invariant provided  $V(\phi)$ and $U(\phi)$ are of the
form  
\begin{equation}
V(\phi) = f_{1}  e^{\alpha\phi},  U(\phi) =  f_{2}
e^{2\alpha\phi}
\end{equation}
and $\varphi_{a}$ is transformed according to
\begin{equation}
\varphi_{a}   \rightarrow   \lambda_{a} \varphi_{a}  
\end{equation}
(no sum on a) which means
\begin{equation}
\Phi \rightarrow \biggl(\prod_{a} {\lambda}_{a}\biggr) \Phi \\ \equiv \lambda 
\Phi	 \end{equation}
such that
\begin{equation} 
\lambda = e^{\theta}
\end{equation}	
and	 
\begin{equation}
\phi \rightarrow \phi - \frac{\theta}{\alpha}.     	
\end{equation}

In this case we call the scalar field $\phi$ needed to implement scale 
invariance "dilaton".

\section{The conserved dilatation current}

Since there is the symmetry (12), (14), (15), (16), (17), according to 
Noether's theorem, there 
is a conserved current given by (since the variation of the lagrangian 
density vanishes under such symmetry),

\begin{equation}
j^{\mu} = \frac {\partial L} {\partial (\partial_{\mu} \varphi_{a})} \delta \varphi_{a}
+ \frac {\partial L} {\partial (\partial_{\mu} \phi)} \delta \phi  
\end{equation} 

since in the first order formalism 
$ \frac {\partial L} {\partial (\partial_{\mu} g_{\alpha \beta})} = 0 $ and 
$\delta \Gamma^{\lambda}_{\mu\nu}= 0 $ under the symmetry (12), (14), (15), 
(16), (17).

Let us now consider what we should take for $\delta \varphi_{a}$. 
As part of the dilatation symmetry, we have that 
$\varphi_{a} \rightarrow \lambda_{a} \varphi_{a}$ (no sum on a) and since
$\biggl(\prod_{a} {\lambda}_{a}\biggr) \equiv \lambda= e^{\theta}$, we have,
taking a transformation infinitesimally close to the identity, i.e.
${\lambda}_{a} = 1 + {\epsilon}_{a}$, with ${\epsilon}_{a} << 1$ and 
all ${\epsilon}_{a}$ equal, so that ${\epsilon}_{a} = {\theta}/4$ and since
also $\delta \phi = - \frac {\theta} {\alpha} $, that the conserved dilatation 
current is,

\begin{equation}
j^{\mu}_{\theta} = -\frac {\theta}{\alpha} \Phi \partial^{\mu} \phi +
\theta \varepsilon^{\mu\nu\alpha\beta} \varepsilon_{abcd} \varphi_{a}
\partial_{\nu} \varphi_{b} \partial_{\alpha}           
\varphi_{c} \partial_{\beta} \varphi_{d} L_{1}  \equiv \theta j^{\mu}_{D}
\end{equation} 

Notice that in the derivation of the conserved current (19) we have taken 
a very particular choice of the parameters ${\lambda}_{a}$ (all of them equal).
One can ask the question: what if we relax this assumption?, noticing that
one can retain the scale symmetry even when we add to the transformations
of the $\varphi_{a}$ transformations where the volume $\Phi$ is not 
transformed.

Indeed one can study the effect of the volume preserving diffeomorphisms$^{6,7}$

\begin{equation}
\varphi_{a}^{'} = \varphi_{a}^{'} (\varphi_{a})
\end{equation}
such that 
\begin{equation}
 \varepsilon_{abcd} \frac { \partial \varphi_{a}^{'}}{ \partial \varphi_{f}}                                               
 \frac { \partial \varphi_{b}^{'}}{ \partial \varphi_{g}}
 \frac { \partial \varphi_{c}^{'}}{ \partial \varphi_{h}}
 \frac { \partial \varphi_{d}^{'}}{ \partial \varphi_{i}} = \varepsilon_{fghi}             
\end{equation}

In this case, it has been shown that$^{7}$ the "composite gauge field"
(which enters in (19))                                                 
\begin{equation}
A_{\mu \nu \alpha} = \frac {1}{4} \varphi_{a} \varepsilon_{abcd}
\frac { \partial \varphi_{b}}{ \partial x^{\mu}}
\frac { \partial \varphi_{c}}{ \partial x^{\nu}} 
\frac { \partial \varphi_{d}}{ \partial x^{\alpha}} 
\end{equation}

transforms under a diffeomorphism (21) as
\begin{equation}
A_{\mu \nu \alpha} \rightarrow A_{\mu \nu \alpha} + \partial_{[\mu} 
\Lambda_{\nu \alpha]}
\end{equation} 

and in the case the transformation is infinitesimal
\begin{equation}
\varphi_{a}^{'} = \varphi_{a} + \Gamma_{a} (\varphi_{b}), 
\frac {\partial \Gamma_{a}}{ \partial \varphi_{a} } = 0
\end{equation}

the divergence free condition in $\varphi_{a}$ space for $\Gamma_{a}$ 
implies that  
 
\begin{equation}
\Gamma_{a} (\varphi_{b}) = \frac {1}{2} \varepsilon_{abcd}
\frac {\partial \Gamma_{cd}}{ \partial \varphi_{b} }
\end{equation}

for some $\Gamma_{cd}$ and $\Lambda_{\nu \alpha} $ in (23) is then given by$^{7}$

\begin{equation}
\Lambda_{\nu \alpha} = 
[ ( 1 - \frac {1}{4} \varphi_{b} \frac {\partial}{ \partial \varphi_{b}}) 
\Gamma_{cd} - 
\frac {3}{2} \varphi_{b}
\frac { \partial }{ \partial \varphi_{[c}} \Gamma_{|b|d]}]
\frac { \partial \varphi_{c}}{ \partial x^{\nu}}
\frac { \partial \varphi_{d}}{ \partial x^{\alpha}}
\end{equation}
 
 In this case, it is a straightforward problem to show that if
we perform a volume preserving diffeomorphism in the internal 
scalar space, to the original dilatation current we add a piece that is 
separately conserved (one has to use eq. (27) of next section for
this elementary proof). 
  
\section{The Equations of Motion}

	We will now work out the equations of motion for arbitrary choice
of $V(\phi)$ and $U(\phi)$. We study afterwards the choice (15) which
allows us to
obtain the results for the scale invariant case and also to see what
differentiates this from the choice of arbitrary $U(\phi)$ and  $V(\phi)$ 
in a very
special way.

	Let us begin by considering the equations which are obtained from
the variation of the fields that appear in the measure, i.e. the
$\varphi_{a}$
fields. We obtain then  
\begin{equation}		
A^{\mu}_{a} \partial_{\mu} L_{1} = 0   	
\end{equation}
where  $A^{\mu}_{a} = \varepsilon^{\mu\nu\alpha\beta}
\varepsilon_{abcd} \partial_{\nu} \varphi_{b} \partial_{\alpha}
\varphi_{c} \partial_{\beta} \varphi_{d}$. Since it is easy to
check that  $A^{\mu}_{a} \partial_{\mu} \varphi_{a^{\prime}} =
\frac{\delta aa^{\prime}}{4} \Phi$, it follows that 
det $(A^{\mu}_{a}) =\frac{4^{-4}}{4!} \Phi^{3} \neq 0$ if $\Phi\neq 0$.
Therefore if $\Phi\neq 0$ we obtain that $\partial_{\mu} L_{1} = 0$,
 or that
\begin{equation}
L_{1} = \frac{-1}{\kappa} R(\Gamma,g) + \frac{1}{2} g^{\mu\nu}
\partial_{\mu} \phi \partial_{\nu} \phi - V = M	 
\end{equation}
where M is constant.

	Let us study now the equations obtained from the variation of the
connections $\Gamma^{\lambda}_{\mu\nu}$.  We obtain then
\begin{equation}
-\Gamma^{\lambda}_{\mu\nu} -\Gamma^{\alpha}_{\beta\mu}
g^{\beta\lambda} g_{\alpha\nu}  + \delta^{\lambda}_{\nu}
\Gamma^{\alpha}_{\mu\alpha} + \delta^{\lambda}_{\mu}
g^{\alpha\beta} \Gamma^{\gamma}_{\alpha\beta}
g_{\gamma\nu}\\ - g_{\alpha\nu} \partial_{\mu} g^{\alpha\lambda}
+ \delta^{\lambda}_{\mu} g_{\alpha\nu} \partial_{\beta}
g^{\alpha\beta}         \\
 - \delta^{\lambda}_{\nu} \frac{\Phi,_\mu}{\Phi}
+ \delta^{\lambda}_{\mu} \frac{\Phi,_\nu}{\Phi} =  0	
\end{equation}
If we define $\Sigma^{\lambda}_{\mu\nu}$    as
$\Sigma^{\lambda}_{\mu\nu} =
\Gamma^{\lambda}_{\mu\nu} -\{^{\lambda}_{\mu\nu}\}$
where $\{^{\lambda}_{\mu\nu}\}$   is the Christoffel symbol, we
obtain for $\Sigma^{\lambda}_{\mu\nu}$ the equation 
\begin{equation}
	-  \sigma, _{\lambda} g_{\mu\nu} + \sigma, _{\mu}
g_{\nu\lambda} - g_{\nu\alpha} \Sigma^{\alpha}_{\lambda\mu}
-g_{\mu\alpha} \Sigma^{\alpha}_{\nu \lambda}
+ g_{\mu\nu} \Sigma^{\alpha}_{\lambda\alpha} +
g_{\nu\lambda} g_{\alpha\mu} g^{\beta\gamma} \Sigma^{\alpha}_{\beta\gamma}
= 0 
\end{equation}		 
where  $\sigma = ln \chi, \chi \equiv \frac{\Phi}{\sqrt{-g}}$.
      	
	The general solution of (30) is 
\begin{equation}
\Sigma^{\alpha}_{\mu\nu} = \delta^{\alpha}_{\mu}
\lambda,_{\nu} + \frac{1}{2} (\sigma,_{\mu} \delta^{\alpha}_{\nu} -
\sigma,_{\beta} g_{\mu\nu} g^{\alpha\beta})
\end{equation}
where $\lambda$ is an arbitrary function due to the $\lambda$ - symmetry
of the
curvature$^{8}$  $R^{\lambda}_{\mu\nu\alpha} (\Gamma)$,
\begin{equation}
\Gamma^{\alpha}_{\mu\nu} \rightarrow \Gamma^{\prime \alpha}_{\mu\nu}
 = \Gamma^{\alpha}_{\mu\nu} + \delta^{\alpha}_{\mu}
Z,_{\nu}
\end{equation} 
Z  being any scalar (which means $\lambda \rightarrow \lambda + Z$).
  
	If we choose the gauge $\lambda = \frac{\sigma}{2}$, we obtain
\begin{equation}
\Sigma^{\alpha}_{\mu\nu} (\sigma) = \frac{1}{2} (\delta^{\alpha}_{\mu}
\sigma,_{\nu} +
 \delta^{\alpha}_{\nu} \sigma,_{\mu} - \sigma,_{\beta}
g_{\mu\nu} g^{\alpha\beta}).
\end{equation}

	Considering now the variation with respect to $g^{\mu\nu}$, we
obtain
\begin{equation}	 	
\Phi (\frac{-1}{\kappa} R_{\mu\nu} (\Gamma) + \frac{1}{2} \phi,_{\mu}
\phi,_{\nu}) - \frac{1}{2} \sqrt{-g} U(\phi) g_{\mu\nu} = 0
\end{equation}
Solving for $R = g^{\mu\nu} R_{\mu\nu} (\Gamma)$  and introducing in
(28), we obtain a constraint,
\begin{equation}
M + V(\phi) - \frac{2U(\varphi)}{\chi} = 0
\end{equation}
that allows us to solve for $\chi$,
\begin{equation}
\chi = \frac{2U(\phi)}{M+V(\phi)}.
\end{equation}

	To get the physical content of the theory, it is convenient to go
to the Einstein conformal frame where 
\begin{equation}
\overline{g}_{\mu\nu} = \chi g_{\mu\nu}		    
\end{equation}
and $\chi$  given by (36). In terms of $\overline{g}_{\mu\nu}$   the non
Riemannian contribution $\Sigma^{\alpha}_{\mu\nu}$
disappears from the equations, which can be written then in the Einstein
form ($R_{\mu\nu} (\overline{g}_{\alpha\beta})$ =  usual Ricci tensor)
\begin{equation}
R_{\mu\nu} (\overline{g}_{\alpha\beta}) - \frac{1}{2} 
\overline{g}_{\mu\nu}
R(\overline{g}_{\alpha\beta}) = \frac{\kappa}{2} T^{eff}_{\mu\nu}
(\phi)	 	
\end{equation}
where
\begin{equation}	 
T^{eff}_{\mu\nu} (\phi) = \phi_{,\mu} \phi_{,\nu} - \frac{1}{2} \overline
{g}_{\mu\nu} \phi_{,\alpha} \phi_{,\beta} \overline{g}^{\alpha\beta}
+ \overline{g}_{\mu\nu} V_{eff} (\phi)
\end{equation}
and 	
\begin{equation}
V_{eff} (\phi) = \frac{1}{4U(\phi)}  (V+M)^{2}.
\end{equation}
	
	In terms of the metric $\overline{g}^{\alpha\beta}$ , the equation
of motion of the Scalar
field $\phi$ takes the standard General - Relativity form
\begin{equation}
\frac{1}{\sqrt{-\overline{g}}} \partial_{\mu} (\overline{g}^{\mu\nu} 
\sqrt{-\overline{g}} \partial_{\nu}
\phi) + V^{\prime}_{eff} (\phi) = 0.
\end{equation} 

	Notice that if  $V + M = 0,  V_{eff}  =  0$ and $V^{\prime}_{eff} 
= 0$ also, provided $V^{\prime}$
is finite and $U \neq 0$ and regular there. This means the zero cosmological 
constant state is achieved without any sort of fine tuning. This is the basic
feature that characterizes theories with the additional measure $\Phi$
where $L_{1}$ and $L_{2}$ are $\varphi_{a}$ independent and allows them to solve the
cosmological constant problem$^{4}$. It should be noticed that the equations
of motion in terms of  $\overline{g}_{\mu\nu}$ are perfectly regular at 
$V + M = 0$ although the transformation (37) is singular at this point. In terms
of the original metric $g_{\mu\nu}$ the equations do have a singularity at 
$V + M = 0$. The existence of the singular behavior in the original frame implies
the vanishing of the vacuum energy for the true vacuum state in the bar frame,
but without any singularities there.

	In what follows we will study (40) for the special case of global
scale invariance, which as we will see displays additional very special
features which makes it attractive in the context of cosmology.

	Notice that in terms of the variables $\phi$,
$\overline{g}_{\mu\nu}$, the "scale"
transformation becomes only a shift in the scalar field $\phi$, since
$\overline{g}_{\mu\nu}$ is
invariant (since $\chi \rightarrow \lambda^{-1} \chi$  and $g_{\mu\nu}
\rightarrow \lambda g_{\mu\nu}$)
\begin{equation}
\overline{g}_{\mu\nu} \rightarrow \overline{g}_{\mu\nu}, \phi \rightarrow
\phi - \frac{\theta}{\alpha}.
\end{equation}

 If $V(\phi) = f_{1} e^{\alpha\phi}$  and  $U(\phi) = f_{2}           
e^{2\alpha\phi}$ as                                                           
required by scale                                                             
invariance (14), (16), (17), (18), (19), we obtain from (40)                  
\begin{equation}                                                              
        V_{eff}  = \frac{1}{4f_{2}}  (f_{1}  +  M e^{-\alpha\phi})^{2}        
\end{equation}                                                                
                                                                              
        Since we can always perform the transformation $\phi \rightarrow      
- \phi$ we can                                                                
choose by convention $\alpha > O$. We then see that as $\phi \rightarrow      
\infty, V_{eff} \rightarrow \frac{f_{1}^{2}}{4f_{2}} =$ const.                
providing an infinite flat region. For the interpretation of this flat
region of the potential in terms of the restoration of a conserved charge
and therefore the possibility of interpreting it as the appearance of a 
Goldstone boson, see the discussion section.
A  minimum of this effective potential is achieved at zero without fine tuning         
 for any case where $\frac{f_{1}}{M} < O$  

\section{The Equation of Motion of the Scalar Field from the Conservation
Law}

We can derive the equations of motion for $\phi$ directly from the 
variation with respect to $\phi$, this gives eq.(41). It is interesting to
see however that from the local conservation 
$ \partial _{\mu} j^{\mu}_{D} = 0 $, with $j^{\mu}_{D}$ given by (19)
gives exactly the same equation. Indeed, demanding 
that $ \partial _{\mu} j^{\mu}_{D} = 0 $, with $j^{\mu}_{D}$ given by 
(19) implies,
\begin{equation}
\Phi L_{1} - \frac {1}{\alpha} \partial_{\mu} (\Phi g^{\mu \nu} \partial_{\nu} \phi) = 0
\end{equation}

but $\Phi = \chi \sqrt {-g} =  \chi ^{-1} \sqrt {-\overline{g}} $
and $g^{\mu \nu} \Phi = g^{\mu \nu} \chi \sqrt {-g} = 
\overline{g}^{\mu \nu}\sqrt {-\overline{g}} $, so that (44) becomes
(using that $L_{1} = M$) that,

\begin{equation}
-\alpha \chi ^{-1} M +
 \frac {1}{\sqrt {-\overline{g}}} \partial_{\mu} (\overline{g}^{\mu \nu}
\sqrt {-\overline{g}} \phi) = 0
\end{equation}

but for the form (43) of the effective potential it is satisfied that
$ V_{eff}^{'} = -\alpha \chi ^{-1} M $, so that the current conservation
is nothing but the scalar field equation when the choice (13) is made.

\section{ The infrared behavior of the dilaton current and the reason
for Goldstone's theorem breakdown}

To see the basic reasons why the dilatation current (19) has an infrared 
singular behavior, let us consider the spatial behavior of the 
$\varphi_{a}$ fields for the case of a simple spatially flat Robertson-Walker
solution of the form
\begin{equation}
ds^{2} = -dt^{2} + R^{2}(t) (dx^{2} +dy^{2} + dz^{2}), 
 \phi = \phi(t)
\end{equation}
 From eq.(35) and (46), we see also that $\chi = \chi(t)$. Then, since
$\chi = \chi(t) = \frac {\Phi}{R^{3}(t)}$, we get that,
\begin{equation}
\Phi = R^{3}(t) \chi(t) = \varepsilon^{\mu\nu\alpha\beta}  \varepsilon_{abcd}                  
\partial_{\mu} \varphi_{a} \partial_{\nu} \varphi_{b} \partial_{\alpha}      
\varphi_{c} \partial_{\beta} \varphi_{d}   
\end{equation}

(47) can be solved by taking 
\begin{equation} 
\varphi_{1} = x, \varphi_{2} = y, \varphi_{3} = z,
\varphi_{4} = - \frac{1}{4!} \int \chi(t^{'}) R^{3}(t^{'}) dt^{'}
\end{equation}

For this case, with a time dependent scalar field  $\phi(t)$ and with 
$\varphi_{a}$ given by (48), the spatial components of the current 
$j^{\mu}_{D}$, as given by (19) diverge linearly as $x^{i} \rightarrow \infty$
($x^{1}=x, x^{2}=y, x^{3}=z$). In fact 
$j^{i}_{D} \rightarrow M x^{i} \chi(t) R^{3}(t)$ as 
$x^{i} \rightarrow \infty $  

Such current does indeed give flux at infinity. The current grows linearly with
distance, so that the total flux is proportional to the volume enclosed
and obviously the total dilatation charge is not conserved here.

In the context of theories with additional measure $\Phi$, there are other
instances where Goldstone's theorem can fail. For example in (7)-(11), 
take the 
model, without scale invariance where $U(\phi) = \Lambda = $ constant and
$V(\phi) = J\phi$, the model has a symmetry up to the integral of a total 
divergence, 
$ \phi \rightarrow \phi + c,  c =$ constant. In this case, since 
$V_{eff}  = \frac{1}{4 \Lambda}  (J \phi + M)^{2}$, we see again that no 
Goldstone boson is present in the particle spectrum, as was observed
in the last paper of Ref.4, although the conserved currents were not
obtained there. Working out the conserved
current associated with this symmetry, we see that it is
$j^{\mu}_{shift}=  \Phi \partial^{\mu} \phi +         
J \varepsilon^{\mu\nu\alpha\beta} \varepsilon_{abcd} \varphi_{a}         
\partial_{\nu} \varphi_{b} \partial_{\alpha}                                  
\varphi_{c} \partial_{\beta} \varphi_{d} $  which has a singular infrared
behavior, exactly for the same reasons the dilatation current has
(i.e. because of the singular behavior of the $\varphi_{a}$ fields at
spatial infinity).

Notice that the potential $V_{eff}  = \frac{1}{4 \Lambda}  (J \phi + M)^{2}$
is contained in the class of potentials being discussed here, i.e.
$V_{eff}  = \frac{1}{4f_{2}}  (f_{1}  +  M e^{-\alpha\phi})^{2}$, for the 
limit $\alpha \rightarrow 0, \alpha M \rightarrow$ constant, 
 $\alpha^{2} M, \alpha^{3} M,... \rightarrow 0 $, so that 
$V_{eff}  = \frac{1}{4f_{2}}  (f_{1}  +  M - M \alpha \phi)^{2}$ ,
so that if $f_{1}  +  M$ is kept fixed in this limit, we obtain a 
purely quadratic potential. The flat region has, in this limit been pushed
out and has gone away.

\section{Discussion and Conclusions}
We have studied the structure of the conserved quantities in the scale
invariant theories where an additional measure $\Phi$ is introduced.

In this case, we have seen that although the global scale invariance
plays a crucial role in determining the structure of the theory, allowing 
only a very specific type of effective potential for a scalar field, 
there are however no Goldstone Bosons associated with the spontaneous
breaking of this symmetry.

The reason for this is that although a locally conserved dilatation current 
can be defined, the spatial part of the current has a singular infrared 
behavior, making it impossible then to prove that the dilatation charge 
will be conserved. As we have seen this also happens in other global 
shift-like symmetries, which can be understood as singular limits of the
scale symmetries discussed in the body of the paper.

An interesting phenomena that takes place in these scale invariant models
is that as long as we go to a regime where the constant of integration $M$
can be ignored, a dilatation symmetry charge appears to be conserved and 
a flat potential, which means an associated Goldstone boson appears in this
regime.

In fact, we see that the infrared singular part of the current (19) is that 
proportional to $M$ (recall that $L_{1}=M$) and if $M$ can be ignored for some reason and set to 
zero, such infrared problems go away. In terms of the effective potential 
$V_{eff}$, we see that as long as $M$ can be ignored, for $\alpha \phi 
\rightarrow \infty$, $V_{eff}$ becomes a flat potential and $\phi$ becomes
in this regime a true Goldstone boson.

We see therefore that potentials with flat regions, as required in the 
cosmological models of new inflation$^{9, 10}$ for example can appear from a 
first principle, scale invariance. Such flat potentials may be of interest
also in the present stages of the Universe$^{5}$.

For the discussion we have limited ourselves to the simplest case where we
have discussed only a scalar field. The introduction of scale invariant
masses for other fields does not change in any way the qualitative features
like the singular infrared behavior and associated non conservation of the
scale charge which is the focus of this paper.

 \section{Acknowledgments}

I would like to thank J. Bekenstein, A. Davidson, A.Guth,  A. Kaganovich,  
P.Mannheim, E.Nissimov, S.Pacheva and L.C.R. Wijewardhana for 
conversations on the subjects discussed here. In particular, I thank 
P.Mannheim for strongly encouraging me to look at the reasons why 
Goldstone's theorem can fail for some symmetries in theories with 
the additional measure $\Phi$
discussed here and for discussions on scale invariance, a demand we both 
agree has to play a fundamental role in gravity, but where he has 
developed a different approach$^{11}$ to the one explained here.

\break


\begin{thebibliography}{99}
\bibitem{1} Y.Nambu and G. Jona-Lasinio, Phys. Rev. 122 (1961) 345;
 Phys. Rev. 124 (1961) 246; J.Goldstone, Nuovo Cimento 19 (1961) 15.
\bibitem{2} See for example  S.Coleman's Erice lectures "Dilatations",
reprinted in S.Coleman, "Aspects of Symmetry", Cambridge University 
Press (1985).
\bibitem{3} G.t'Hooft, Phys. Rev. Lett. 37 (1976) 8; Phys. Rev. D14 
(1976) 3432.
\bibitem{4} E.I. Guendelman and A.B. Kaganovich, Phys. Rev.,  D53,
(1996)
7020; E.I. Guendelman and A.B. Kaganovich, Proceedings of the third
Alexander Friedmann International Seminar on Gravitation and Cosmology,
ed. by Yu. N. Gneding, A.A. Grib and V.M. Mostepanenko (Friedmann
Laboratory Publishing, St. Petersburg, 1995);
        E.I. Guendelman and A.B. Kaganovich,  Phys. Rev., D55, (1997)
5970;
        E.I. Guendelman and A.B. Kaganovich, Mod. Phys. Lett, A12, (1997)
2421;
 E.I. Guendelman and A.B. Kaganovich, Phys. Rev., D56, (1997) 3548;
        E.I. Guendelman and A.B. Kaganovich, Hadronic Journal, 21, (1998)
19;
        E.I. Guendelman and A.B. Kaganovich, Mod. Phys. Lett., A13, (1998)
1583;
        F. Gronwald, U. Muench and F.W. Hehl, Hadronic Journal, 21, (1998)
3;
        E.I. Guendelman and A.B. Kaganovich, Phys. Rev., D57,  (1998)
7200;
        E.I. Guendelman and A.B. Kaganovich, "Gravity Cosmology and 
Particle Field Dynamics without the Cosmological Constant Problem", to
appear in the Proceedings of the sixth International Symposium on
Particle, Strings and Cosmology, PASCOS-98;
 E.I. Guendelman, "Gauge Condensates and Gauge Dynamics, the cosmological 
and strong CP problems", to appear in the Int. Journ. of Mod. Phys. A.;
        E.I. Guendelman and A.B. Kaganovich, "Field Theory Models without
the Cosmological Constant problem", Plenary talk (given by E.I.
Guendelman) at the fourth Alexander Friedmann International Seminar on
Gravitation and Cosmology, gr-qc/9809052 and extended version of this,
gr-qc/9905029, to appear in Phys. Rev. D15.
\bibitem{5} E.I. Guendelman, gr-qc/9901017, to appear in Mod. Phys. Lett. A;
 E.I. Guendelman, gr-qc/9901067.
\bibitem{6} E.I.Guendelman, E.Nissimov and S.Pacheva, Phys. Lett. B360 (1995)
57; C.Castro, Int. Journ. of Mod. Phys. A13 (1998) 1263. 
\bibitem{7}E.I.Guendelman, E.Nissimov and S.Pacheva, hep-th/9903245.
\bibitem{8} A. Einstein, "The Meaning of Relativity", MJF books, NY (1956), 
see Appendix II.
\bibitem{9} For a non technical review and a good collection of
further
references on different aspects of inflation see A. Guth, "The
Inflationary Universe", Vintage, Random House (1998). For a more technical
review see E.W. Kolb and M.S. Turner, "The Early Universe", Addison Wesley
(1990).
\bibitem{10} The original papers on new inflation are A.D. Linde, 
Phys. Lett,  108B, (1982) 389; 	A. Albrecht and P.J. Steinhardt, 
Phys. Rev. Lett, 48,  (1982) 1220.
\bibitem{11} see for example, P.Mannheim, gr-qc/9903005 and references.
\end{thebibliography}
\end{document}